\documentclass{MITcsail}
\usepackage{indentfirst}

\title{Molecular dynamics predictions of transport properties for carbon dioxide hydrates under pre-nucleation conditions using TIP4P/Ice water and EPM2, TraPPE, and Zhang carbon dioxide potentials}

\newcommand{\angstrom}{\mbox{\normalfont\AA}}
\author
{Andr\'e Guerra\orcidA{}\footnote{Correspondence E-mail: andre.guerra@mail.mcgill.ca}, Samuel Mathews\orcidB{}, Jennifer Tram Su, Milan Mari\'c\orcidC{}, Phillip Servio\orcidD{}, Alejandro D. Rey\orcidE{}\footnote{Correspondence E-mail: alejandro.rey@mcgill.ca}\\
\vspace{1em} 
\normalfont{\small Department of Chemical Engineering, McGill University, Montr\'eal, QC, Canada}\\
}

\begin{document}

\maketitle
\thispagestyle{firstpagestyle} 

\begin{abstract}
(1) Introduction: New technologies that leverage gas hydrates phenomena include carbon capture and sequestrations. These processes are often semi-continuous and require regulation of the system's flow properties for proper operation. Accurate computational models for the viscosity of carbon dioxide hydrate systems at pre-nucleation conditions can be important for process design and control of such technologies. (2) Methods: This work validates the viscosity predictions of molecular dynamics simulations using previously measured experimental data. The TIP4P/Ice force field was used to model water, while the EPM2, TraPPE, and Zhang force fields were used for carbon dioxide. The Green-Kubo and Einstein formulations of viscosity and diffusivity were used in this work. (3) Results: All force fields overpredicted viscosity when compared to experimental data, but EPM2 resulted in lower discrepancies. Additionally, EPM2 was determined to model molecular behavior expected from the macroscopic trends in viscosity with respect to temperature and pressure. (4) Conclusions: The EPM2 force field more accurately predicted the viscosity of carbon dioxide hydrates systems at pre-nucleation conditions relative to TraPPE and Zhang.
\end{abstract}

\section{Introduction}
Among the Arctic permafrost and the subsea sediments of Earth’s continental margins are ice-like crystalline solids\cite{Bergeron2010}. What separates these compounds from normal forms of ice (e.g., hexagonal) are gas inclusions - guest gas molecules that reside within the cavities formed by the lattice structure. Known as gas hydrates, these non-stoichiometric compounds form when gaseous molecules and water freeze at high pressures and low temperatures. Depending on these conditions, as well as the identity of the gas molecule, hydrate properties may vary, and structurally take on one of three main forms: sI, sII, or sH, which differ in size, structure, and number of occupancies\cite{Mathews2022}. Water molecules maintain the hydrate structure through hydrogen bonds, while the entrapped molecules provide stability through weak van der Waals interactions\cite{Bergeron2010}.

In the past two decades, new technologies that utilize the formation of gas hydrates to accomplish a process have been proposed. Some examples include pre- and post-combustion carbon capture\cite{Aaron2005,Linga2007,Kang2000a}, gas separations\cite{Eslamimanesh2012a,Fan2009a}, transport and storage of natural gas\cite{Gudmundsson1994,Mimachi2015}, and water desalination\cite{Park2011}. These technologies mostly leverage the high gas-to-solid volume ratio of gas hydrates (up to 184:1), and the hydrate cages size selectivity to guest species to accomplish their process\cite{Sloan2008,Eslamimanesh2012a}. Due to the semi-continuous nature of these new technologies and the absence of oil emulsions in their systems, recent studies have explored the dynamic viscosity of methane and carbon dioxide hydrate systems in aqueous systems\cite{Guerra2022baseline,McElligott2022_nanotube}. Rheological studies of these systems require highly specialized and expensive equipment to be conducted, in addition to difficult experimental conditions to be achieved. As a result, the use of computational methods to predict transport properties, like dynamic viscosity and diffusivity, is desirable.

Computational methods such as Density Functional Theory (DFT) and Molecular Dynamics (MD) can provide insight into molecular structure and transport phenomena that are otherwise difficult, if not impossible, to obtain experimentally. DFT is well suited for atomic-scale quantum calculations to examine a material's static properties, while MD is preferred for nanoscale dynamic processes such as transport properties. However, results from computational methods are best interpreted and utilized in combination with experimental data. Our research group has developed an integrated experimental-computational platform for investigating the material sciences of gas hydrate systems\cite{Guerra2022_review}. This has led to contributions in interfacial phenomena\cite{Mirzaeifard2018,Mirzaeifard2019} and mechanical and transport properties of gas hydrate systems\cite{Vlasic2019IRspectra,Vlasic2016,Vlasic2017,Vlasic2019,Zhu2022piezoelasticity,Zhu2022atomisticcontinuumcharac,Guerra2022_md_methane,Zhu_frac_mech_2023,Zhu_2023_TOEC}. In this work, we implement the platform to study the transport properties of carbon dioxide hydrate systems as predicted by molecular dynamics. We use experimental data to validate the dynamic viscosity predictions by our models. The experimental data is presented elsewhere\cite{Guerra2022baseline}.

Molecular dynamics simulations implement molecular trajectories by integrating Newton’s equations of motion. Atomic interactions are mapped by force fields, which are parametrized models that represent the potential energy of an atomic system as a function of the individual atomic positions\cite{Rapaport2004}. This potential energy arises from the sum of bonded and non-bonded interactions. Bonded forces describe how covalent bonds behave, including changes in bond length, bond angle, and bond torsion along dihedrals. These are often modeled as harmonic oscillators. Non-bonded forces arise between non-covalently bonded atoms, including electrostatic Coulombic interactions and repulsive van der Waals forces, modeled by Coulomb’s law and the Lennard-Jones 12/6 potential model, respectively. For simplicity, these N-body interactions are approximated as a pairwise additive model. By tracking how an atomic system evolves over time, MD simulations rely on statistical mechanical principles of the ensemble and the ergodic hypothesis to predict macroscopic material properties, such as viscosity and diffusivity, from nanoscale interactions. 

Molecular simulations have been used to investigate gas hydrate systems including the nucleation\cite{Jimenez-Angeles2014_GH_nucleation_md,Liang2010_GH_crystal_sim,Hawtin2008_GH_nucleation_MD,Moon2003_GH_formation_MD,Zhang2008_GH_formation_MD,Walsh2009_GH_nucleation_MD,Zhang2015_GH_nucletaion_MD,Guo2013_GH_nucleation_MD,Yan2005_GH_formation_MD} and dissociation\cite{English2005_GH_decomp_MD,Baez1994,English2013_GH_decomp_MD,Ding2007_GH_decomp_MD,Myshakin2009_GH_decomp_MD,Iwai2010_GH_decomp_MD,Smirnov2012_GH_decomp_MD} of hydrates, the mobility of guest species between cages\cite{Geng2009_GH_mobility_MD,Tung2011_GH_mobility_MD,QI2011_GH_mobility_MD,Bai2012_GH_mobility_MD}, surface effects on nucleation\cite{Bagherzadeh2012_GH_surf_MD,Liang2011_GH_surf_MD,Bai2011_GH_surf_MD,Bai2012_GH_surf_MD,Bai2015_GH_surf_MD}, and nucleation inhibition mechanisms by poly(vinyl pyrrolidone) and poly(vinyl caprolactam)\cite{Li2017_MD_KHI,Kuznetsova2012_GH_inhibPVCap_MD,Maddah2018_GH_inhib_MD,Xu2016}. A recent review by Qi et al.\cite{Qi2021_GH_MD_review} expands on recent advances in gas hydrate research accomplished by the use of molecular simulations. The transport properties of gas hydrate systems have not been explored to the same extent. Recently, our research has conducted molecular simulations to model methane hydrate systems and to predict their dynamic viscosity\cite{Guerra2022_md_methane}. These predictions were compared to experimental results for model validation. This previous work has demonstrated the TIP4P/Ice water model to improve the dynamic viscosity prediction of subcooled water over the widely used TIP4P/2005\cite{Guerra2022_md_methane}. 

The scope of this study is to examine the performance of equilibrium molecular dynamics viscosity predictions of carbon dioxide hydrate system at pre-nucleation conditions. This is achieved through comparison with experimental data and by conducting various hydrogen bond analyses. In this work, we use the TIP4P/Ice force field potential to model water molecules, and we test the performance of three widely accepted carbon dioxide force field potentials EPM2\cite{Harris1995}, TraPPE\cite{Potoff2001}, and Zhang\cite{Zhang2005}. The Green-Kubo and Einstein formulations are used on a rigorously equilibrated large molecular system (5472-9072 atoms) to predict their dynamic viscosity and diffusivity, respectively. Additionally, the Stokes-Einstein formulation for viscosity is introduced and its performance is evaluated in this context.

\section{Tools and Methods}
\subsection{Software Packages}

Molecular dynamics (MD) simulations were conducted using the Large-scale Atomic/Molecular Massively Parallel Simulator (LAMMPS) package, an open-source code developed and maintained by Sandia National Laboratory and Temple University\cite{Thompson2022}. Atom trajectories are dynamically tracked via the integration of Newton’s equations of motion coupled with varying interatomic potentials. When analyzed, these interacting particles serve as a predictive model for macroscopic material properties. MD simulations in LAMMPS require an initial configuration of atoms and molecules. These configurations were developed using the PACKMOL\cite{Martinez2009} and Moltemplate\cite{Jewett2021} packages. PACKMOL is a packing optimization algorithm that places atoms according to spatial constraints set by the user (in this work, a separation tolerance of 2.5\angstrom). As opposed to the standard population commands in LAMMPS, these constraints give rise to structural complexity while reducing repulsive inter-molecular forces, effectively lessening the computational power required during the initial part of the simulation process. Moltemplate is a cross-platform molecule builder that prepares MD systems by assigning atoms and molecules their corresponding parameters according to the desired force field potential. Atomic mass, charges, and molecule bond and angle parameters associated with the Optimizing Potentials for Liquid Simulations OPLS) All-Atom TIP4P/Ice force field were assigned for water, while parameters associated with the Elementary Physical Model (EPM2), Transferable Potential for Phase Equilibria (TraPPE), and Zhang force fields were assigned for carbon dioxide. Finally, MDAnalysis is a python library developed to handle and analyze molecular trajectories, which was used by this work to conduct hydrogen bond analyses.

\subsection{Simulation Design}
\subsubsection{Force Field Potentials}
The Optimizing Potentials for Liquid Simulations All-Atom (OPLS-AA) force field was used to model all water molecules. Compared to the coarse-grained version united-atom (UA) form, the all-atom form represents hydrogen atoms explicitly, making it better suited for simulating gas hydrate systems due to the high presence of hydrogen bonding\cite{Jorgensen1983,Jorgensen1988,Jorgensen1996}. In particular, the TIP4P/Ice four-site water model was specifically designed for estimating properties of water in its solid-state form\cite{Abascal2005_ice}. A recent study considering methane hydrates demonstrated that this model had improved performance at predicting transport properties of water at low temperatures over the TIP4P/2005 model\cite{Guerra2022_md_methane}.

Carbon dioxide was modeled using the three-site models TraPPE\cite{Potoff2001}, Zhang\cite{Zhang2005}, and EPM2\cite{Harris1995}. The Zhang model is considered the best all-around performing force field potential for predicting the transport properties of pure carbon dioxide, including thermodynamic property predictions\cite{Aimoli2014}. EPM2 and TraPPE were the next best-performing force field potentials. The Zhang model parameters were optimized to accurately predict liquid volumetric properties and phase-equilibria\cite{Zhang2005}. The EPM2 model is an improvement on the rigid EPM model by introducing a flexible bond angle potential, which more accurately predicts the liquid-vapor coexistence curve for pure carbon dioxide\cite{Harris1995}. Finally, the TraPPE model was developed to expand the applicability of pure-component models to multi-component mixtures involving n-alkanes\cite{Potoff2001}.

\subsubsection{LAMMPS Input}
The TIP4P/Ice water model was used to simulate all systems. The largest of the systems considered in this work consisted of 2976 molecules, which is one order of magnitude greater than the classic system of 256 molecules that suffers from finite size effects\cite{MonterodeHijes2018}. The systems were defined by the \textit{full} atom style, the \textit{lj/cut/tip4p/long} pairstyle with 12\angstrom~OM site and coulombic cut-off lengths, and the \textit{pppm/tip4p} kspace space style with a dimensionless relative force accuracy of $1 \times 10^{-4}$. Bond and angle styles were modeled as \textit{harmonic}, while interatomic interactions between dissimilar non-bonded atoms were calculated through the Lorenz-Berthelot arithmetic mixing rules. Three molecular representations of carbon dioxide were used. The TraPPE and Zhang models (rigid C=O bond angle) and EPM2 model (semi-flexible C=O bond angle). Water molecule bonds and angles were kept rigid via the \textit{shake} command. This command is not recommended to constrain angles at 180 degrees as it results in numerical solver difficulties\cite{Thompson2022}. As a linear molecule, carbon dioxide molecules modeled by TraPPE and Zhang had their bonds and angles kept rigid via the \textit{rigid} commands, which treat all atoms in a molecule as one moving body. All rigid models had their bond and angle harmonic constants set to 1000 $kcal/mol/rad^2$ to ensure rigidity during the minimization step. To advance the molecular trajectories, Newton’s equations of motion were numerically integrated using the Velocity Verlet algorithm with a 2-femtosecond timestep. Table~\ref{tbl:ff_params} details the list of parameters for the TIP4P/Ice, EPM2, TraPPE, and Zhang force fields.

\begin{table}[H]
  \caption{Molecular force field parameters implemented in LAMMPS in this work.}
  \label{tbl:ff_params}
  \newcolumntype{C}{>{\centering\arraybackslash}X}
\begin{tabularx}{\textwidth}{CCCCC}
    \hline
\textbf{Attribute, Units} & \textbf{Zhang \cite{Zhang2005}} & \textbf{TraPPE \cite{Potoff2001}} & \textbf{EPM2 \cite{Harris1995}} &\textbf{TIP4P/Ice \cite{Abascal2005_ice}} \\
    \hline
O mass, g/mol               & 15.999            & 15.999            & 15.999                & 15.9994 \\
H mass, g/mol               & -                 & -                 & -                     & 1.008 \\
C mass, g/mol               & 12.011            & 12.011            & 12.011                & - \\
\hline
O charge, e                 & $-$0.2944         & $-$0.35           & $-$0.3256             & $-$1.1794 \\
H charge, e                 & -                 & -                 & -                     & 0.5897 \\
C charge, e                 & 0.5888            & 0.7               & 0.6512                & - \\
\hline
OH bond $r_o$, \angstrom    & -                 & -                 & -                     & 0.9572 \\
CO bond $r_o$, \angstrom    & 1.163             & 1.1672            & 1.149                 & - \\
\hline
OCO angle $\theta$          & $180^{\circ}$     & $180^{\circ}$     & $180^{\circ}$         & - \\
HOH angle $\theta$          & -                 & -                 & -                     & $104.52^{\circ}$ \\
OM distance, \angstrom      & -                 & -                 & -                     & 0.1577 \\
\hline
H-H LJ $\epsilon$, kcal/mol & -                 & -                 & -                     & 0 \\
C-C LJ $\epsilon$, kcal/mol & 0.057131          & 0.053477          & 0.055713              & - \\
O-O LJ $\epsilon$, kcal/mol & 0.163711          & 0.15647           & 0.159455              & 0.21084 \\
\hline
C-C LJ $\sigma$, \angstrom  & 2.7918            & 2.8               & 2.757                 & - \\
O-O LJ $\sigma$, \angstrom  & 3.0               & 3.05              & 3.033                 & 3.1668 \\
H-H LJ $\sigma$, \angstrom  & -                 & -                 & -                     & 0 \\
$r_c$, \angstrom            & 12                & 12                & 12                    & 12 \\
\hline
\end{tabularx}
\noindent{\footnotesize{\textbf{Notes}: H: hydrogen, O: oxygen, C: carbon, O-O: oxygen-oxygen interactions, H-H: hydrogen-hydrogen interactions, charge units in multiples of an electron charge (e), M is the massless fourth site in the TIP4P water model, $r_c$: cutoff distance of Coulombic and Lennard-Jones (LJ) interactions,~$\sigma$: distance for zero potential energy in LJ potential,~$\epsilon$ is the depth of the LJ potential well.}}
\end{table}

\subsubsection{Carbon Dioxide Hydrate Systems}
The pre-nucleation gas hydrate systems defined in this study consisted of carbon dioxide and water. For a binary mixture of two phases (liquid-vapour), the Gibbs phase rule imposes two degrees of freedom. In this study, they were taken up by temperature and pressure, leaving the carbon dioxide liquid concentration fixed at every condition. Carbon dioxide concentrations in water at low temperatures can be described in an analogous way as methane systems were examined by Lekvam and Bishnoi \cite{Lekvam1997,Krichevsky1935,Anderson2005a,Carroll1991_co2_solubility}. Here, a modified version of Henry’s law for higher pressures is used. Henry’s law has been used to describe the liquid solubility of carbon dioxide in water for hydrate systems due to the small effect of pressure on solubility at low temperatures\cite{Hashemi2006,Servio2001}. For application in systems at higher pressures, Krichevsky and Kasarnovsky present a modification to Henry's law as shown in Equation~\ref{eqn:KK}\cite{Krichevsky1935}.

\begin{equation}
    \label{eqn:KK}
    ln \left ( \frac{f_2}{x_2} \right ) = ln(H_{2, 1}) + \frac{\bar{v}_2^{\infty}(P - P_1^S)}{RT}
\end{equation}
Where for the gas species, $f_2$ is the fugacity, $x_2$ is its liquid concentration, and $H_{2, 1}$ is Henry's law constant. As well, $P_1^S$ is the partial saturation pressure of the liquid, $\bar{v}_2^{\infty}$ is the partial molar volume of the gas at infinite dilution, $R$ is the gas constant, $P$ is the pressure, and $T$ is the temperature.

The fugacity of carbon dioxide in water is calculated using the Trebble-Bishnoi equation of state (EOS)\cite{Trebble1987}. As this model has been extensively studied, it will not be included here. The partial molar volume, $\bar{v}_2^\infty$, and Henry’s law constant, $H_{2, 1}$, for carbon dioxide are provided by Di et al.\cite{Di2021_co2_solub_model} and Carroll et al.\cite{Carroll1991_co2_solubility}. Each carbon dioxide system consisted of between 1426 to 2976 molecules depending on the liquid concentration calculated. The carbon dioxide concentrations calculated for each condition are presented in Table~\ref{tbl:co2conc}. The calculated concentrations did not vary in pressure within the precision presented. This is in accordance to previous observation that the carbon dioxide concentration in the liquid phase is a weak function of pressure in hydrate systems\cite{Servio2001}. Morevover, the concentration values presented in Table~\ref{tbl:co2conc} are in agreement with experimental and calculated values presented in this previous work\cite{Servio2001}. The molecular system's concentration was produced by packing the simulation box with molecules. However, for the system to remain physically meaningful this work performed an optimization procedure to ensure that the ratio between molecules of each species used achieved the concentration dictated by the description above while still maintaining the target density at the conditions of interest. This created the discrepancy in system size between conditions.

\begin{table}[H]
  \caption{Carbon dioxide concentration in mole fractions for each condition considered in this work.}
  \label{tbl:co2conc}
  \newcolumntype{C}{>{\centering\arraybackslash}X}
\begin{tabularx}{\textwidth}{CCCC}
    \hline
    & \textbf{0 MPag} & \textbf{2 MPag} & \textbf{3 MPag} \\
    & $\pm$0.00005 & $\pm$0.00005 & $\pm$0.00005 \\
    \hline
    \textbf{0 $^{\circ}$C}	& 0.0160	& 0.0160 & 0.0160 \\
    \textbf{4 $^{\circ}$C}	& 0.0189 	& 0.0189 & 0.0189 \\ 
    \textbf{8 $^{\circ}$C}	& 0.0269	& 0.0269 & 0.0269 \\
    \hline
\end{tabularx}
\end{table}

\subsection{Equilibration Procedure}
Accurate transport property estimations require appropriately designed simulations. To this end, Maginn et al.\cite{Maginn2018} present an equilibration procedure based on best practices for transport property estimations from MD simulations. Molecules were randomly assigned an initial velocity based on the Maxwell-Boltzmann distribution at the desired temperature. All systems studied in this work were then equilibrated through a series of simulations in different ensembles. First, the isobaric-isothermal (NPT) ensemble was performed for 25 ns, during which the system’s potential energy and density were monitored. Sufficient equilibration was attained when these parameters stabilized within their expected range\cite{Abascal2005_ice,Orsi2014}. Following this step, the canonical (NVT) ensemble with the Nosé-Hoover thermostat was performed for 50 ns. The system’s density and pressure were monitored throughout the equilibration run to ensure no considerable deviation of the thermodynamic state of the system from the previous NPT step. Following best practices, it is recommended to use the NVT ensemble to estimate transport properties as opposed to the NPT ensemble. In maintaining pressure in an NPT ensemble, volume corrections cause mechanical disturbances to the dynamics of the system. Additionally, properties such as diffusivity and viscosity that are calculated using the Green-Kubo autocorrelation functions, are particularly sensitive to these disturbances due to their dependence on velocity and pressure\cite{Maginn2018}. Furthermore, the temperature and pressure were modulated using damping factors of 100*dt and 1000*dt, respectively, where “dt” is the simulation timestep (2 femtosendons). These were introduced to reduce the disturbances induced by ensemble dynamics.

\subsection{Replicate Production Runs}
In equilibrium molecular dynamics, the simulated systems that are referred to as "production runs" are equilibrated and are the systems from which property calculations can be confidently performed. The ideal ensemble for production runs would be the NVE, in which no barostat or thermostat is implemented. This ensures no disturbances are introduced into the system from these dynamic temperature and/or pressure control algorithms. However, the use of the NVE ensemble is problematic to maintain equilibrium conditions (i.e., maintaining the desired pressure-temperature condition). For the context of transport property estimations, previous work has shown that the estimations of viscosity and diffusivity are indistinguishable between NVE and NVT systems\cite{Fanourgakis2012,Nose1984}. As a consequence of this, this work implements NVT production runs for transport property calculations after the conclusion of the equilibration procedure described above. 

In order to reduce the effect of variability of the autocorrelation functions used to calculate viscosity and diffusivity (described below) on the calculated values, this work implements bootstrap statistical analysis for its production runs. Once the system has been equilibrated as described above, ten ($n=10$) replicate NVT systems were created, each using a different random number generator seed for assignment of the temperature at each pressure-temperature condition. This creates a sample of ten replicate simulations at each desired condition, all of which were simulated in LAMMPS for 1.5 ns for production run calculations to be performed. From the results of the array of systems, a statistical bootstrap with replacement ($N=10,000$) was performed for each pressure-temperature condition. Unless otherwise specified, all transport properties presented in this work are averages from the $N=10,000$ bootstrap procedure. This statistical averaging takes advantage of the inverse proportionality between the calculated values' uncertainty and the number of replicates ($u \propto 1/\sqrt{n}$)\cite{Payal2012}. Moreover, the bootstrap procedure performed here allows for the calculated values to be presented with their standard errors. Finally, the replicate simulations benefit from the computational speed of parallel simulations versus one long simulation, which has been shown to converge to the same value\cite{Zhang2015,Ma2017}.

\subsection{Viscosity and Diffusivity Formulations}
Molecular dynamic simulations rely on the fundamental principles of statistical mechanics, which draw the connection between a system's microstates and its macroscopic material properties\cite{Thompson2022}. Viscosity and diffusivity in particular are studied in this work and can be estimated from fluctuations in pressure and velocity, respectively.  These relations are described by the equilibrium Green-Kubo (GK) time autocorrelation relation and its differential counterpart, the Einstein (Eins) formulation\cite{Green1954,Kubo1957}. Equation~\ref{eqn:vgk} presents the relation between viscosity and the off-diagonal elements of the pressure tensor. In Equation~\ref{eqn:deins}, the Einstein formulation is used to calculate the diffusivity using the mean squared displacement (MSD) of molecules.

\begin{equation}
    \label{eqn:vgk}
    \eta_{GK} = \frac{V}{k_BT}\int_0^\infty\langle P_{\alpha\beta}(t)\cdot P_{\alpha\beta}(0)\rangle dt
\end{equation}

\begin{equation}
    \label{eqn:deins}
    D_{Eins} = \frac{1}{2 d_{\alpha}}\lim_{t\rightarrow\infty}\frac{d}{dt}MSD
\end{equation}
Where $k_B$ is the Boltzmann constant, $d_\alpha$ is the number of dimensions in the simulation ($d_\alpha$ = 3 in this work), $P_{\alpha\beta}$ is an off-diagonal element of the pressure tensor, $MSD \equiv |r_i(t) - r_i(0)|^2$ is the mean squared displacement of molecules, and $r_i$ is the position of the $i^{th}$ molecule.

The Einstein method for diffusivity is often preferred due to its relative stability and shorter time requirement for convergence\cite{Maginn2018}. We use it in this work to establish the Stokes-Einstein formulation for viscosity (Equation~\ref{eqn:vSE}). This formulation utilizes the Stokes-Einstein relation for the diffusivity of spherical particles in low $Re$ flow (Eq.~\ref{eqn:SE}) and the Einstein formulation of diffusivity described above (Eq.~\ref{eqn:deins}). This formulation benefits from the relative stability of the Einstein diffusivity formulation compared to the Green-Kubo viscosity, but it introduces the particle and flow assumptions of the Stokes-Einstein relation (Eq.~\ref{eqn:SE}), which may be poor assumptions here.

\begin{equation}
    D = \frac{k_BT}{6\pi \eta r} \label{eqn:SE}
\end{equation}

\begin{equation}
    \eta_{SE} = \frac{k_BT dt}{\pi r MSD} \label{eqn:vSE}
\end{equation}
Where $MSD \equiv |r_i(t)-r_i(0)|^2$ is the mean squared displacement of molecules, r is molecular radius, and $r_i$ is the position of the \emph{ i}th molecule.

\section{Results and Discussion}
\subsection{Long-Term Equilibrium Indicators}
\begin{figure}[H]
\begin{tabular}{cc}
    \includegraphics[width=0.45\linewidth]{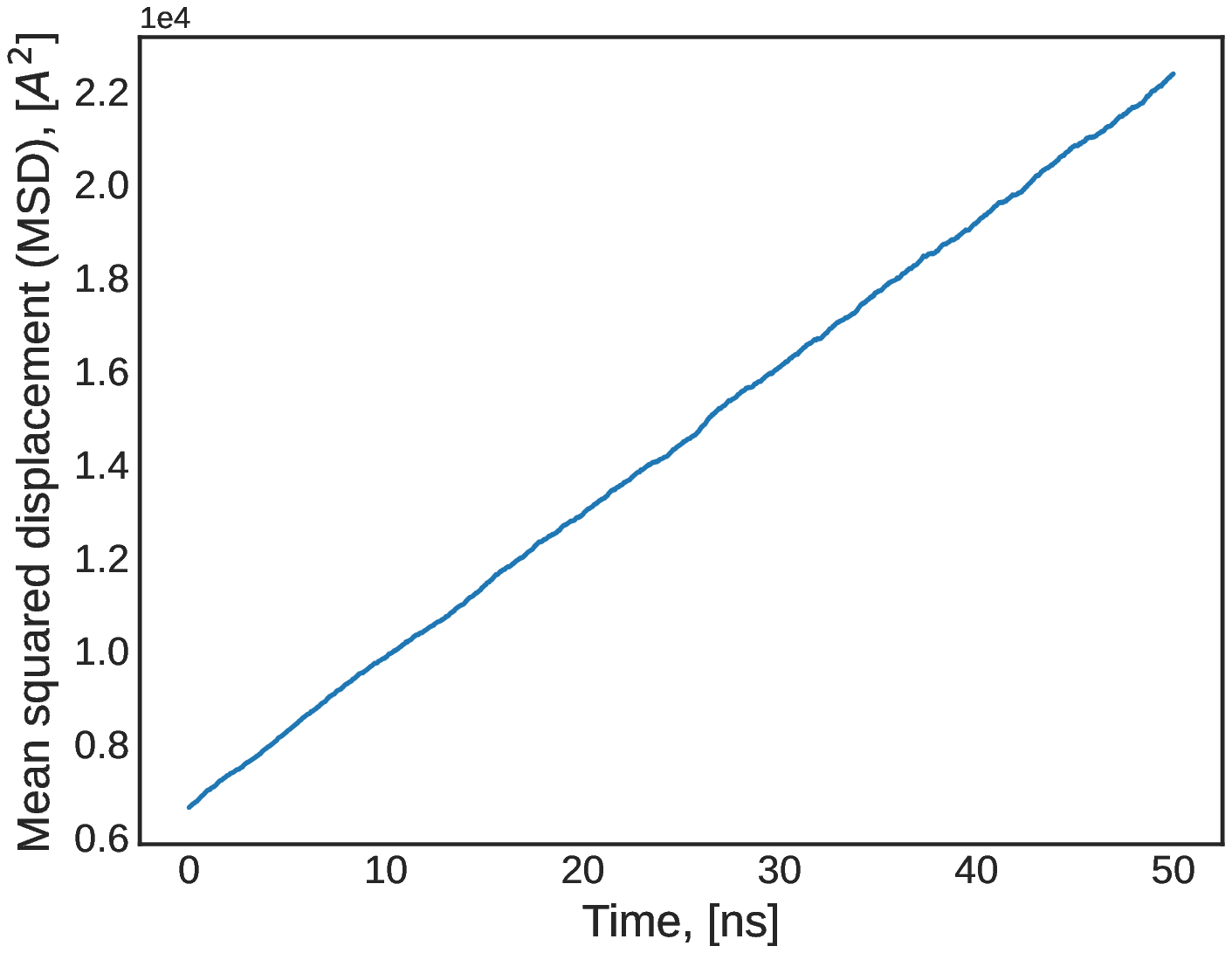}
    &\includegraphics[width=0.45\linewidth]{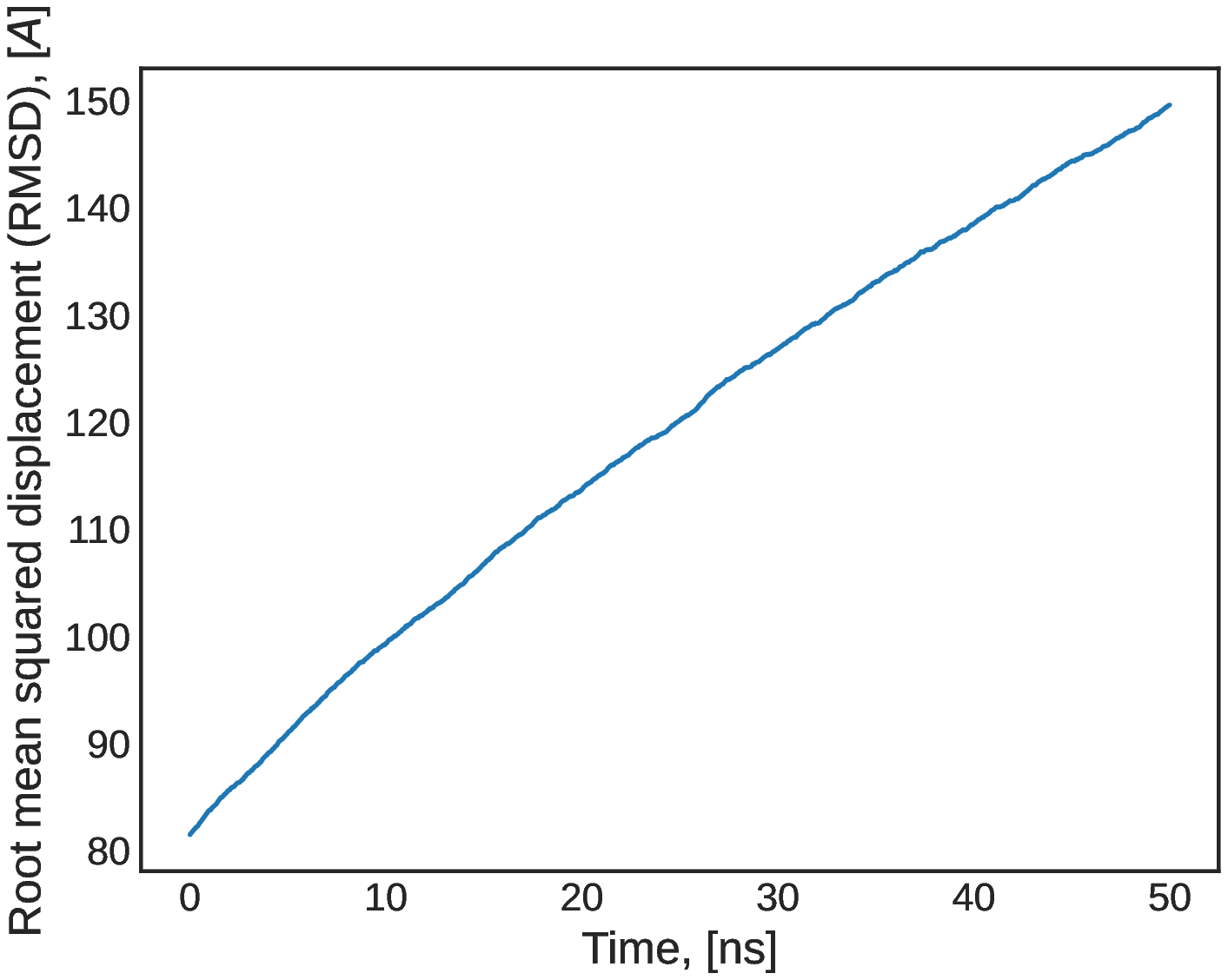}\\
    ({\bf a})&({\bf b})\\
    \includegraphics[width=0.47\linewidth]{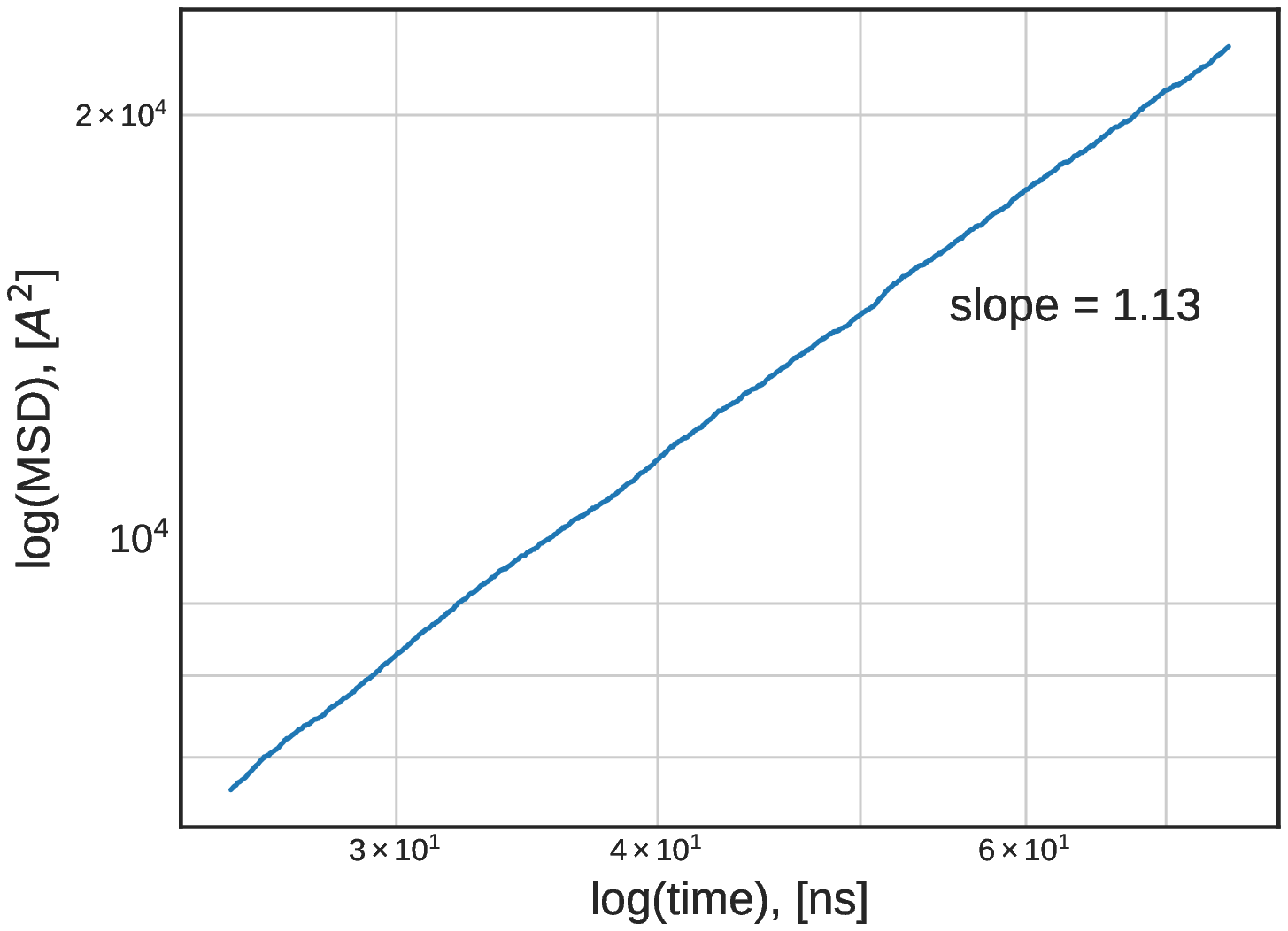}
    & \includegraphics[width=0.47\linewidth]{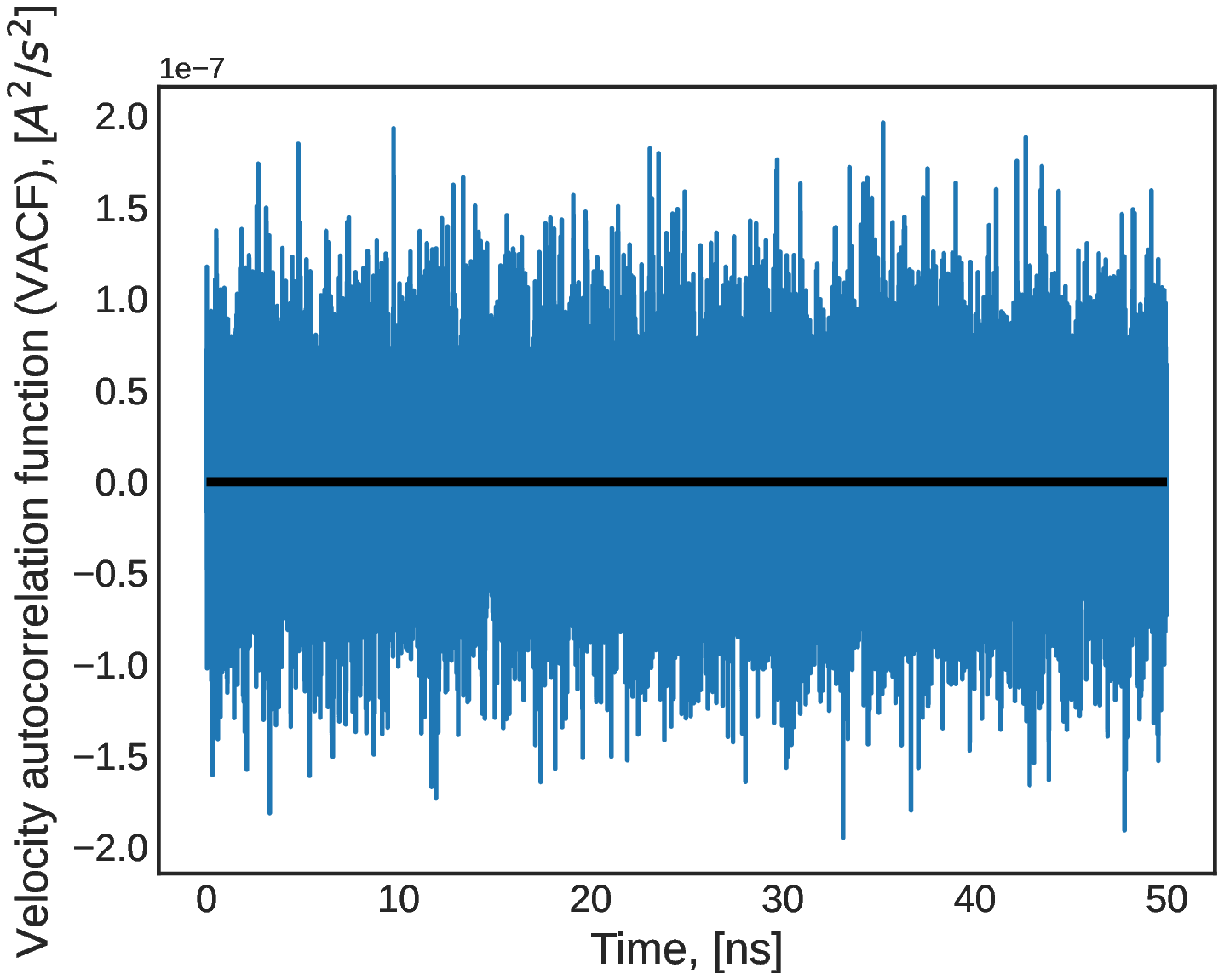}\\
    ({\bf c})   &({\bf d})\\
    \end{tabular}
    \caption{Long-term equilibrium indicators for the TraPPE system at 0 $^{\circ}$C and 0 MPag; (\textbf{a}) mean squared displacement (MSD), (\textbf{b}) root mean squared displacement (RMSD), (\textbf{c}) log-log plot of MSD, and (\textbf{d}) velocity time autocorrelation function (VACF) during the last 50 ns segment of a 100 ns NVT equilibration.}
\label{fig:equil_nvt_ext}
\end{figure}

\begin{figure}[H]
\centering
    \includegraphics[width=0.46\linewidth]{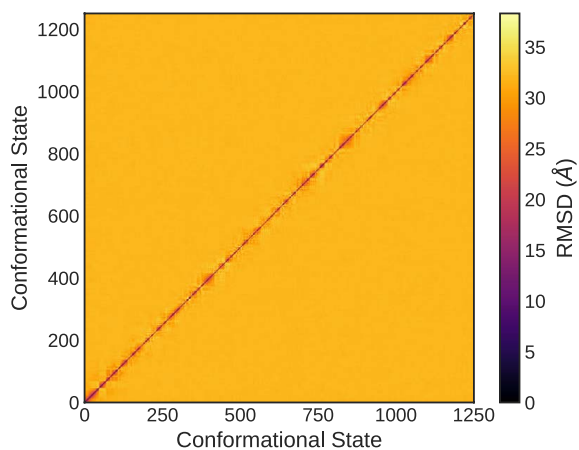}
    \caption{The RMSD self-plot for the TraPPE system at 0 $^{\circ}$C and 0 MPag.}
    \label{fig:equil_rmsd_self}
\end{figure}

The systems simulated in this work have been equilibrated through an extensive procedure, which has been introduced. This procedure entails a systematic approach to equilibration using successive simulations starting with the NPT ensemble and followed by the NVT ensemble and simulating the system for long enough to identify an approach to equilibrium conditions and diffuse regime. This procedure takes into consideration long-term equilibrium indicators including a linear mean squared displacement (MSD) profile (Figure~\ref{fig:equil_nvt_ext}a), a linear root mean squared displacement (RMSD) profile with a final RMSD value that is much larger than the simulation box size (RMSD $>>$ L$_{box}$) (Figure~\ref{fig:equil_nvt_ext}b), a linear log-log plot of MSD vs. time (Figure~\ref{fig:equil_nvt_ext}c), velocity time-autocorrelation that converges to zero (Figure~\ref{fig:equil_nvt_ext}d). Additionally, the molecular trajectories were analyzed through an all-to-all RMSD. This quantifies how far (RMSD) each conformational state of the trajectory has diffused from all others. Figure~\ref{fig:equil_rmsd_self} depicts the RMSD self-plot of a sample system examined. Samples of the long-term equilibrium indicators were presented in Figures~\ref{fig:equil_nvt_ext} and~\ref{fig:equil_rmsd_self}, but these were quantified for all systems considered in this work. Together these indicate a diffuse regime system from which transport properties can be calculated with increased confidence\cite{Maginn2018}. The results presented below were obtained from fully equilibrated systems as identified by the equilibration procedure and the key indicators above. A detailed discussion on the equilibration procedure and equilibrium indicators used in this work were presented elsewhere for similar systems\cite{Guerra2022_md_methane}.

\subsection{Transport Properties}
The dynamic viscosity of the systems simulated in this work was calculated using the Green-Kubo (GK) and Stokes-Einstein (SE) formulations presented above in Equations~\ref{eqn:vgk} and~\ref{eqn:vSE}, respectively. The systems were designed using three different carbon dioxide force fields (EPM2, TraPPE, and Zhang), as described above, and simulated across combinations of three pressures (0, 2, 3 MPag) and temperatures (0, 4, 8 $^{\circ}$C), which span the hydrate-liquid region of the thermodynamic phase diagram of carbon dioxide hydrates. These are conditions that exhibit positive hydrate nucleation pressure driving forces, which in combination with equilibrium molecular dynamics define a pre-nucleation condition. In other words, carbon dioxide hydrates are thermodynamically favored to form under these conditions and theoretically will form if enough time is provided. The timescale of the simulations here is in the order of 75 nanoseconds, which is assumed to be too short for sustained hydrate growth. 

\begin{figure}
\centering
    \begin{tabular}{cc}
        \includegraphics[width=0.46\linewidth]{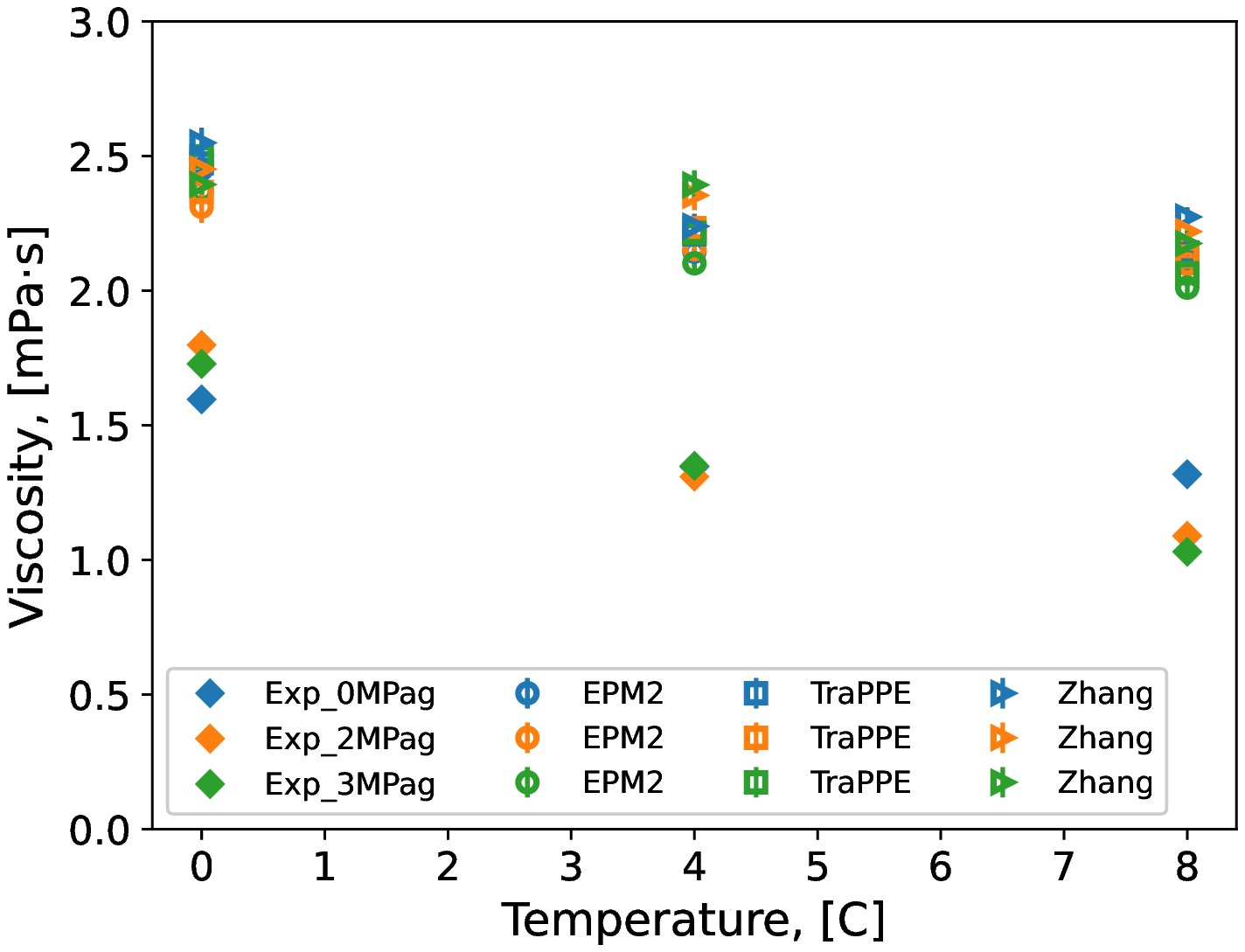}
        &\includegraphics[width=0.47\linewidth]{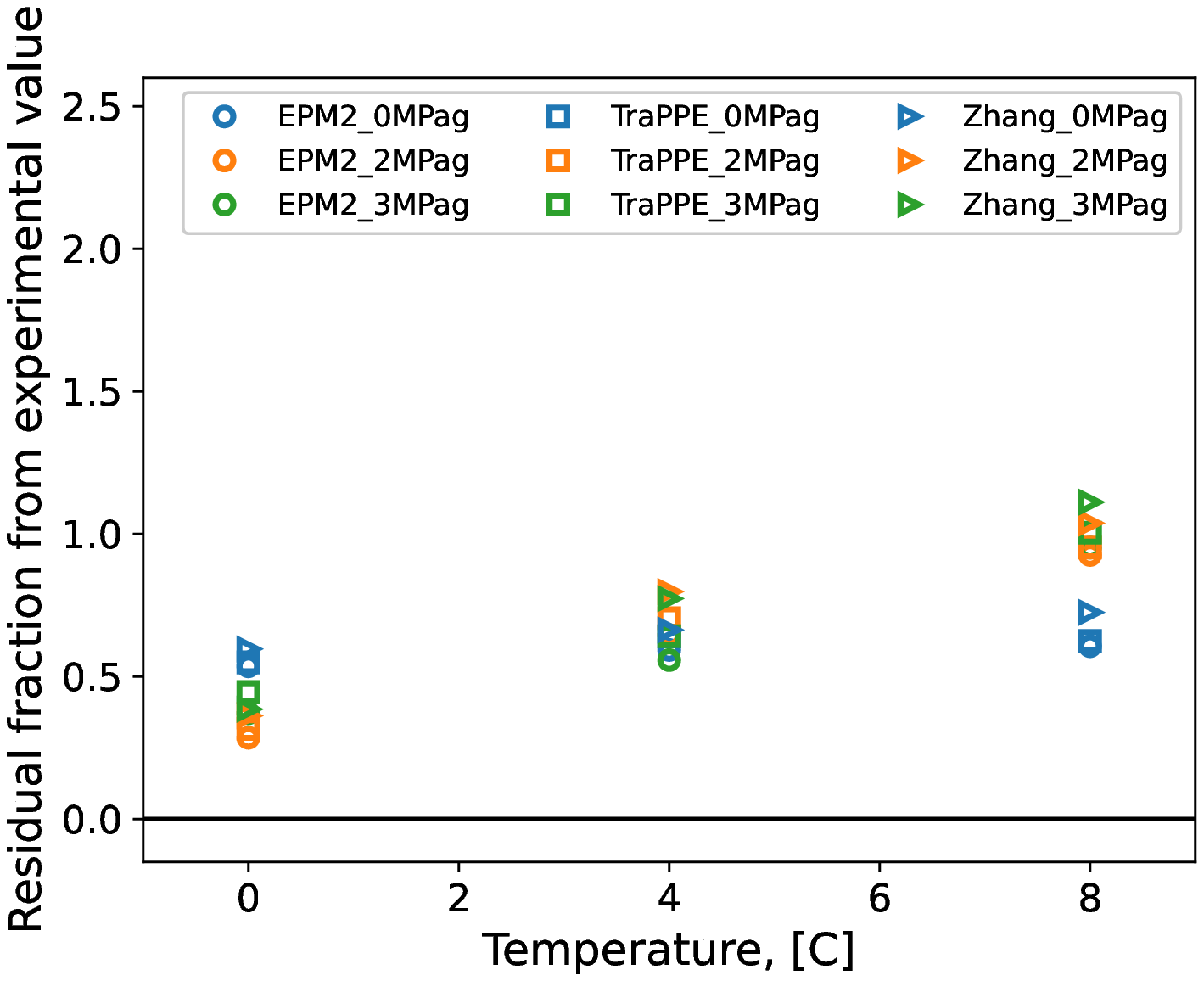}\\
        ({\bf a})&({\bf b})\\
    \end{tabular}
    \caption{The (\textbf{a}) Green-Kubo viscosity of carbon dioxide hydrate systems and (\textbf{b}) the fractional residual difference between simulation and experimental values. Repeated simulations produced samples of each condition's prediction value of size \emph{n} = 10; these samples were bootstrapped with replacement (\emph{N} = 10,000) to calculate mean values which are presented here. Vertical bars are standard errors from bootstrap statistics. Experimental (Exp) values are from previous work presented elsewhere by Guerra et al.\cite{Guerra2022baseline}. Blue: 0 MPag, Orange: 2 MPag, Green: 3 MPag}
    \label{fig:visc_GK}
\end{figure}

Figure~\ref{fig:visc_GK} presents the results from the GK formulation of dynamic viscosity for the conditions and force fields examined in this work. Table~\ref{tbl:visc_perf} presents the average percentage difference between predicted viscosity and experimentally measured values for each force field and viscosity formulation used. In Figure~\ref{fig:visc_GK}(a), the molecular dynamics predictions of viscosity were presented with experimental rheometry data collected elsewhere\cite{Guerra2022baseline}. It is important noting here some limitations associated with the experimental data used in this work. As reported in the original experimental work, the conditions in which gas hydrates ultimately formed (\textit{a.} 0 $^{\circ}$C, 2 MPag, \textit{b.} 0 $^{\circ}$C, 3 MPag and \textit{c.} 2 $^{\circ}$C, 3 MPag) resulted in slightly elevated measured viscosity values. The pre-nucleation viscosity reported was based on a temporal average of the rheometer's measurement, which was elevated by early gas hydrate crystal formation during measurements of the three conditions. The experimental error is not expected to significantly affect the analysis performed in this work, as the predicted viscosity values (Figure~\ref{fig:visc_GK}) suffer from considerably higher error than experimental measurement error.

It is evident that the molecular predictions resulted in overestimated values compared to the experimental data in all cases (conditions and force field combinations). Panel (b) quantifies this discrepancy between predicted and experimental viscosity with the residual fractions (i.e., the percentage difference in fraction format) between these values for all cases. From the residual fractions, it is evident that the EPM2 and TraPPE predictions outperformed the Zhang predictions in most cases. Additionally, the effect of pressure and temperature on the predictions is also apparent. The lowest residual fractions were associated with the atmospheric pressure systems, and the residuals increased with pressure for all force fields examined. Moreover, predictions at zero degrees celsius had a lower variance between force fields. This is likely a result of the phase dynamics at the lowest point in the subcooled water region examined in this work. The system's hydrogen bonding interactions at lower temperatures are enhanced by their approach to phase transition (liquid-to-solid), where hydrogen bonding participates in phase transition. This is observed by higher viscosity predictions and their lower variances.

\begin{table}
  \caption{Average percentage difference between viscosity predicted by the molecular simulations and the experimental measurements.}
  \label{tbl:visc_perf}
  \newcolumntype{C}{>{\centering\arraybackslash}X}
\begin{tabularx}{\textwidth}{CCC}
    \hline
     & \textbf{GK, \%}	& \textbf{SE, \%} \\
     & $\pm$0.05 & $\pm$0.05 \\
    \hline
    \textbf{EPM2}	& 61.0	& 79.0 \\
    \textbf{TraPPE}	& 65.4	& 57.2 \\ 
    \textbf{Zhang}	& 71.7	& 73.2 \\
    \hline
\end{tabularx}
\end{table}

\begin{figure}
\centering
    \begin{tabular}{cc}
        \includegraphics[width=0.46\linewidth]{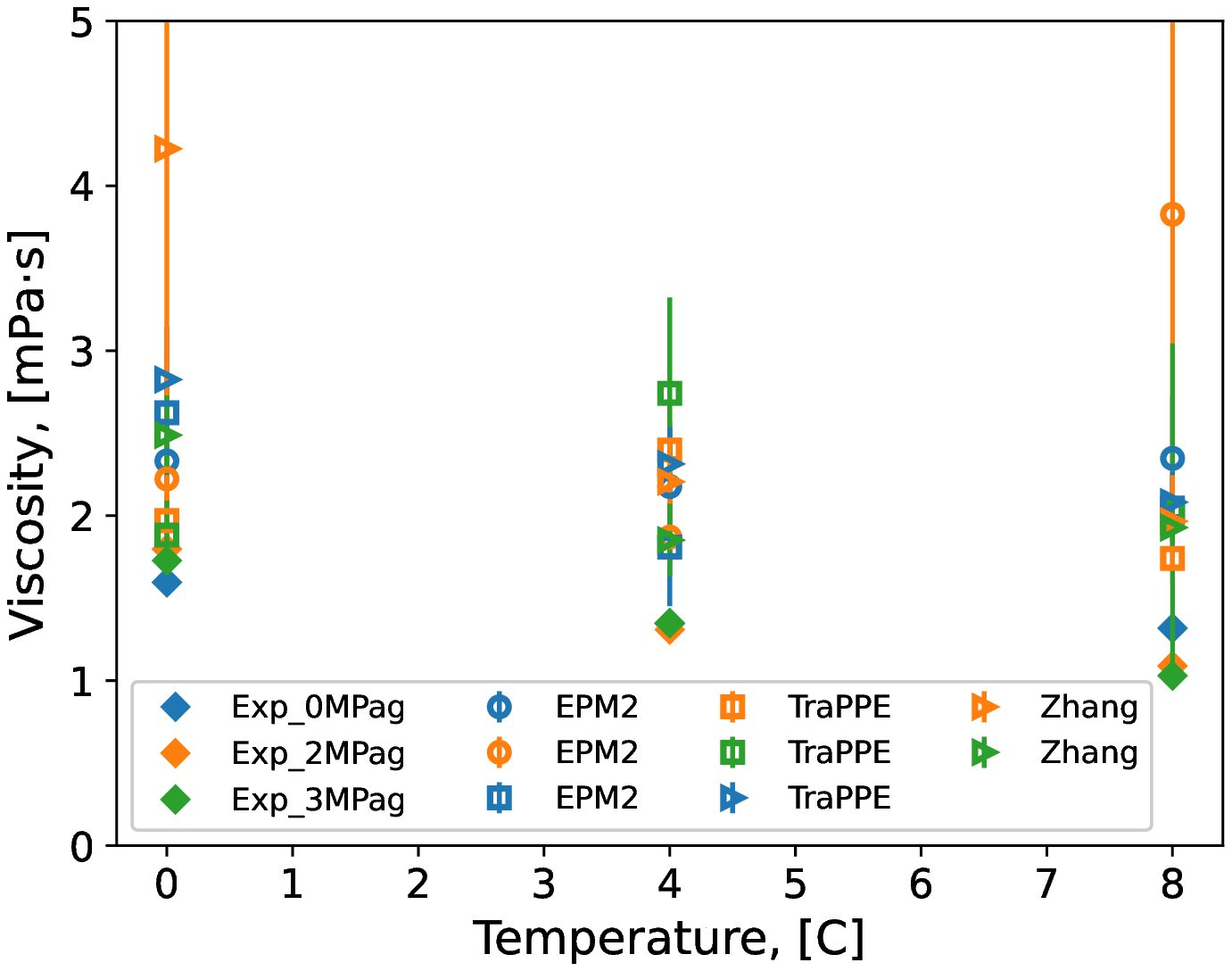}
        &\includegraphics[width=0.47\linewidth]{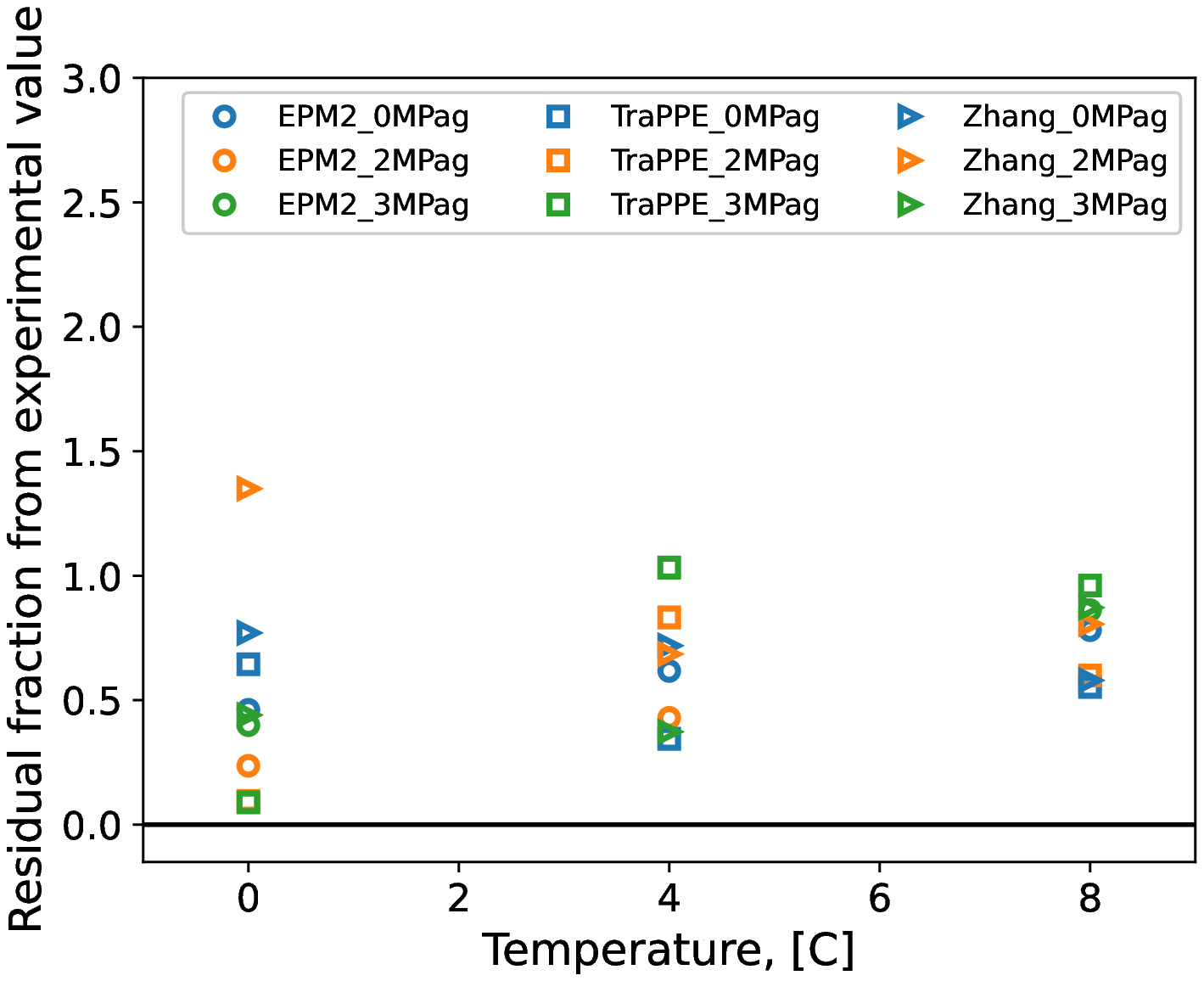}\\
        ({\bf a})&({\bf b})\\
    \end{tabular}
    \caption{The (\textbf{a}) Stokes-Einstein viscosity of carbon dioxide hydrate systems and (\textbf{b}) the fractional residual difference between simulation and experimental values. Repeated simulations produced samples of each condition's prediction value of size \emph{n} = 10; these samples were bootstrapped with replacement (\emph{N} = 10,000) to calculate mean values which are presented here. Vertical bars are standard errors from bootstrap statistics. Experimental (Exp) values are from previous work presented elsewhere by Guerra et al.\cite{Guerra2022baseline}. Blue: 0 MPag, Orange: 2 MPag, Green: 3 MPag}
    \label{fig:visc_SE}
\end{figure}

This work set out to examine the applicability of the SE formulation for one main reason - to attempt to leverage the stability of the commonly used Einstein formulation for diffusivity. The unstable variations in the pressure tensor used in the formulations for GK viscosity (Eq.~\ref{eqn:vgk}) make predictions less reliable than Einstein formulation diffusivity (Eq.~\ref{eqn:deins}), which instead uses the relatively more stable MSD. Ultimately, the formulations for viscosity require a longer simulation time to allow for the stabilization of the pressure tensor elements time-autocorrelations\cite{Maginn2018}. Figure~\ref{fig:visc_SE} presents analogous results as Figure~\ref{fig:visc_GK} above, but the viscosity predictions were obtained from the SE formulation of viscosity. 

Panel (a) in Figure~\ref{fig:visc_SE} presents the dynamic viscosity predictions from each force field examined across the temperature-pressure conditions mentioned above and the experimental viscosity collected by a shear rheometer presented elsewhere\cite{Guerra2022baseline}. As in the case of the GK predictions, the SE viscosity predictions overestimate experimental values in all conditions and all force fields (Table~\ref{tbl:visc_perf}). Panel (b) quantifies the discrepancy between SE predictions and experimental viscosity with the residual fractions between these values. From the residual fractions, the SE formulation of viscosity resulted in an increase in variability of the predictions between the force fields. Although similar trends in temperature and pressure as in the GK predictions are apparent, the performance of the force fields varies across conditions. The EPM2 force field resulted in lower residual fractions than others in most cases. Additionally, the bootstrapped predicted values exhibit greater standard errors (shown as vertical bars) than the GK predictions. This indicates the limitations of the Stokes-Einstein relation used in the SE formulation. The Stokes-Einstein relation relies on a series of assumptions. First, the bodies are assumed spherical, second the drag coefficient of flow over a sphere is used, which imposes the continuum medium assumption. These assumptions are not in agreement with the fundamental aspects of molecular dynamics simulations. This work, however, attempted to determine whether the benefit of the stability of the Einstein diffusivity formulation would out-weight errors associated with the Stokes-Einstein assumption. Provided the results presented in Figure~\ref{fig:visc_SE}, the hypothetical stability advantage of the Stokes-Einstein formulation over the GK prediction is greatly diminished by the inacuracy of the SE assumptions, rendering it inappropriate for this application.

\begin{figure}
\centering
    \includegraphics[width=0.46\linewidth]{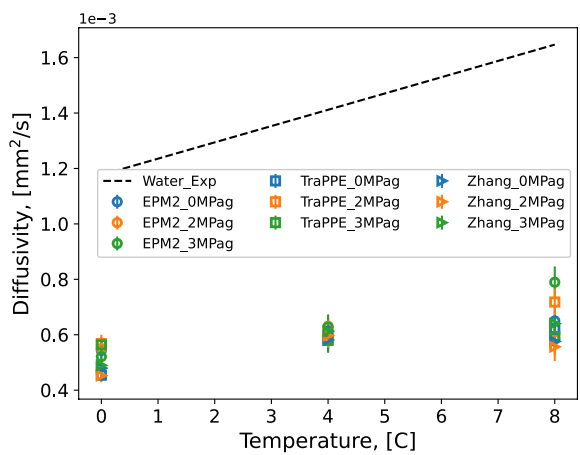}
    \caption{The Einstein diffusivity formulation for carbon dioxide hydrate systems. Repeated simulations produced samples of each condition's prediction value of size \emph{n} = 10; these samples were bootstrapped with replacement (\emph{N} = 10,000) to calculate mean values which are presented here. Vertical bars are standard errors from bootstrap statistics. Exp: linear regression of experimental data for water\cite{Tofts2000,Mills1973,Easteal1989,Harris1980,Gillen1972}. Blue: 0 MPag, Orange: 2 MPag, Green: 3 MPag}
    \label{fig:deins}
\end{figure}

This study also presents in Figure~\ref{fig:deins} the calculated diffusivity of the molecular systems simulated here. Although we are not aware of direct experimental data that would be representative of, and comparable to, the simulated systems here, experimental data for pure water is included in Figure~\ref{fig:deins} to offer a baseline comparison. The positive effect of temperature is evident for all force fields examined, however, their relative performance is uncertain without experimental data. Despite the lack of direct experimental data for validation, the results from the dynamic viscosity analysis above can be used to infer the performance impact on diffusivity predictions due to the inverse proportionality between these two transport properties (Eq.~\ref{eqn:SE}). It is likely that the EPM2 force field and lower pressure conditions provide improved predictions of diffusivity, as they generally did for viscosity.

\subsection{Hydrogen Bond Analyses}
The molecular transport and interactions in aqueous systems are known to be dominated by hydrogen bonds (H-bonds)\cite{Lowry1964}, which are a major contributor to the macroscopic property of dynamic viscosity\cite{Lin2013,Wernet2004,Tokmachev2010}. H-bond librations were recently indirectly measured by quantifying the stretch of OH covalent bonds involved in hydrogen bonding in water through infrared absorption as a relative measurement of viscosity\cite{Ni2019}. Due to the role of H-bonds in the context of transport properties, this work has conducted several H-bond analyses to support the observations discussed above and further quantify the performance of the molecular systems simulated and the effect of carbon dioxide force field choice. In all analyses below, both water and carbon dioxide were considered hydrogen acceptors, and the geometric criteria for a  hydrogen bond were (1) donor-acceptor distance less than 3~\angstrom, and (2) donor-hydrogen-acceptor angle greater than 150$^{\circ}$.

\subsubsection{Hydrogen Bond Structure}
\begin{figure}
\centering
    \includegraphics[width=0.46\linewidth]{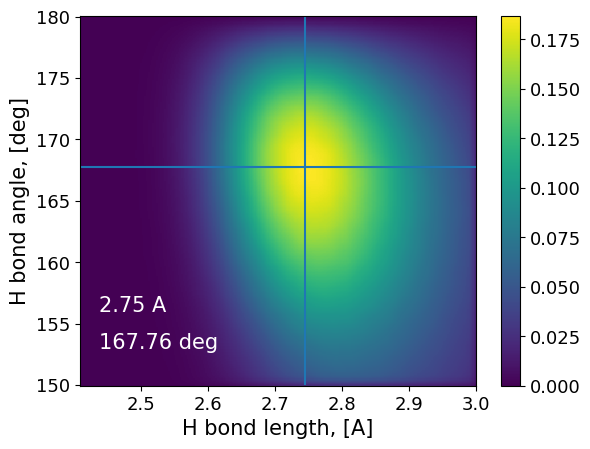}
    \caption{Surface of the probability distribution functions of hydrogen bond length and angle for the EPM2 system simulated at 0$^{\circ}$C and 0 MPag. The most probable values are indicated by the annotation and the vertical and horizontal lines. The colour bar indicates the fractional presence of the H-bond length and angles.}
    \label{fig:hbond_pdf}
\end{figure}

This work conducted Gaussian kernel estimations of the probability density function (PDF) of H-bond length and angle to determine the effect of carbon dioxide force field potentials on the average H-bond structure as a possible source of deviations of viscosity predictions when compared to experimental values. The simulations designed and conducted here resulted in negligible variance in H-bond length and angle. The average H-bond length for all force fields and temperature-pressure conditions was 2.75\angstrom~with a variance of 1.43$\times 10^{-5}$\angstrom. The average H-bond angle was 167.5$^{\circ}$ with a variance of 0.14$^{\circ}$. Figure~\ref{fig:hbond_pdf} presents the PDFs for the EPM2 system simulated at 0$^{\circ}$C and 0 MPag with the most probable bond angle and length indicated by horizontal and vertical lines, respectively. This indicates that on average the simulated H-bond structure, as defined by the bond length and angle, was not affected by force field potential choice nor the temperature-pressure condition. Thus, the variance in viscosity predictions cannot be attributed to H-bond structure effects. The elimination of the H-bond structure as a piezo-viscous driving force is an important result as it significantly reduces the parametric space to be considered in any future reparametrization efforts.

\subsubsection{Hydrogen Bond Lifetime}
\begin{figure}
\centering
    \begin{tabular}{ccc}
        \includegraphics[width=0.3\linewidth]{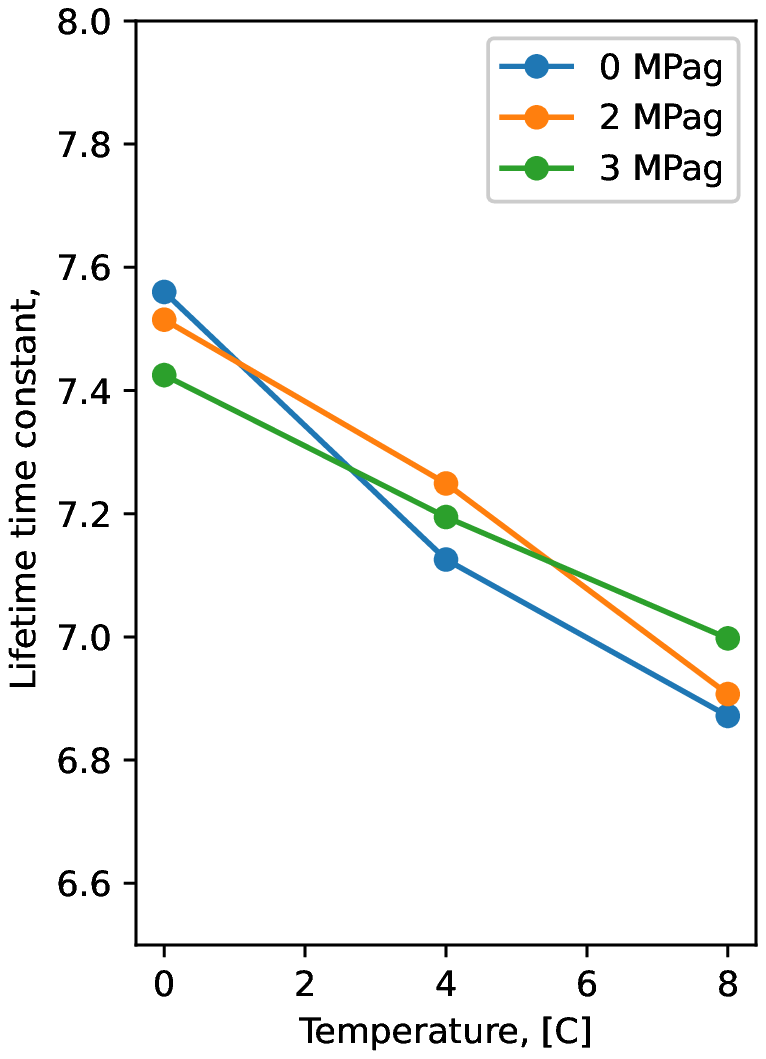}
        &\includegraphics[width=0.3\linewidth]{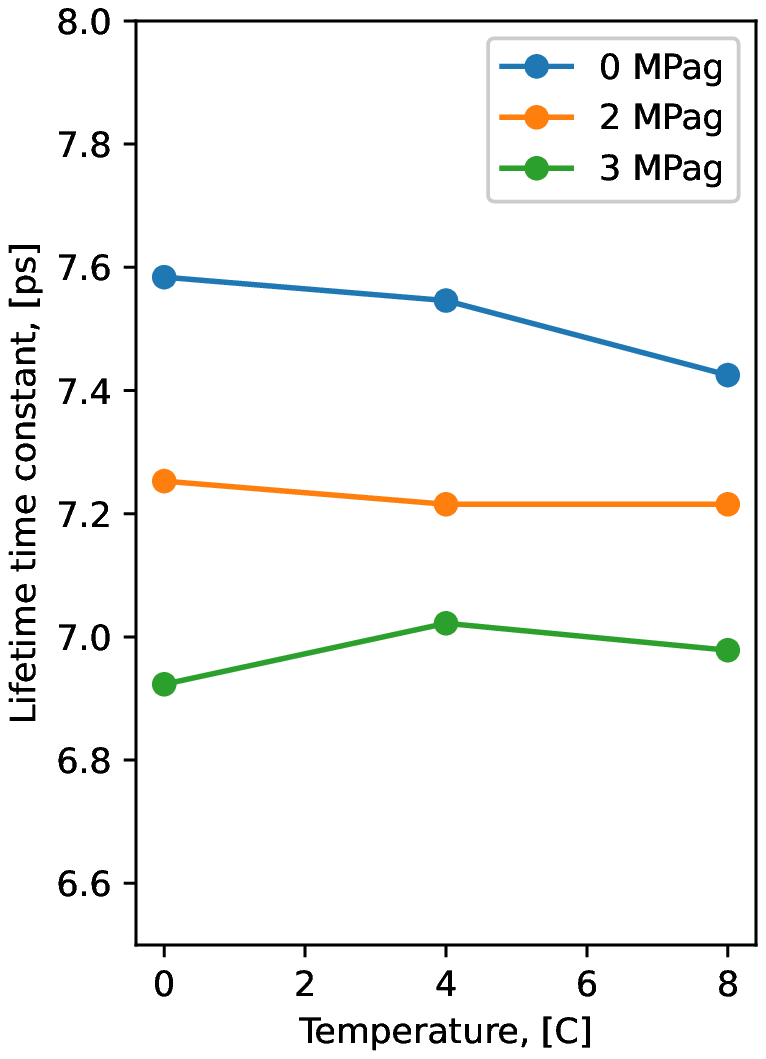}
        &\includegraphics[width=0.3\linewidth]{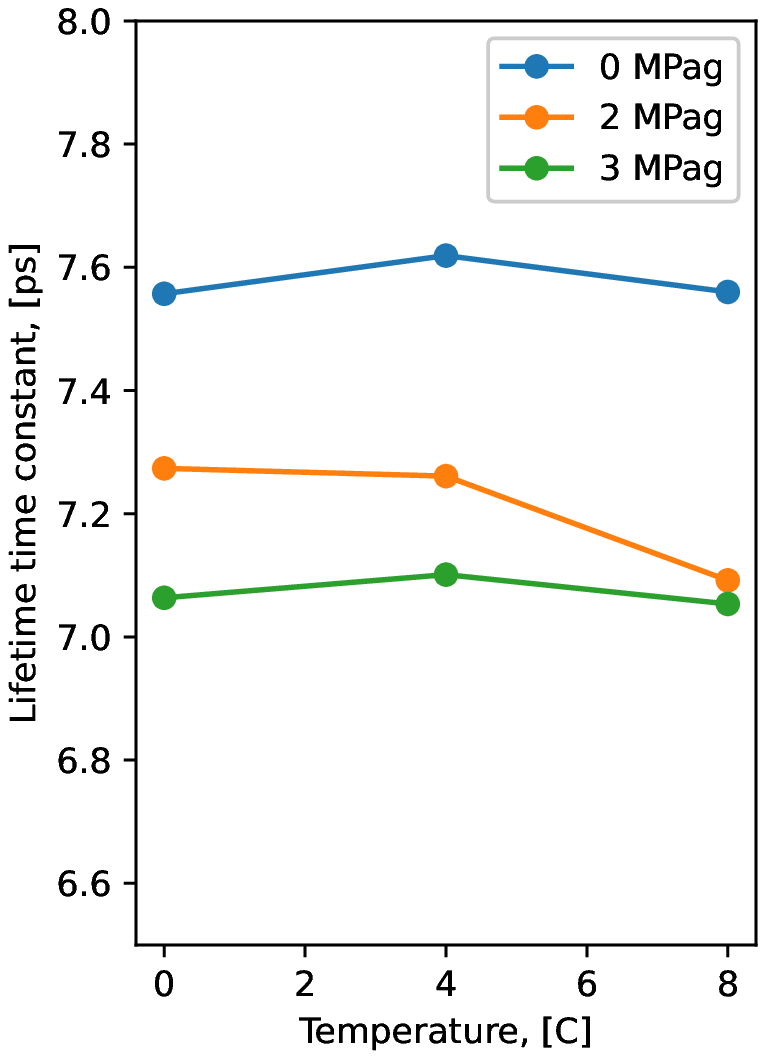}\\
        ({\bf a}) EMP2 &({\bf b}) TraPPE &({\bf c}) Zhang \\
    \end{tabular}
    \caption{The  average lifetimes of hydrogen bonds in each molecular simulation conducted in this work.}
    \label{fig:hbond_lifetime}
\end{figure}

The H-bond lifetime quantifies the average length of time that an H-bond remains intact and is proportional to the system's viscous interactions. It is calculated by the autocorrelation of the binary states (1 or 0) of an H-bond, which indicates whether the H-bond is present (1) or not (0). Detailed descriptions of the algorithm used to quantify H-bond lifetimes have been described elsewhere\cite{Luzar2000_hbond,Gowers2015_hbond}. This work uses the Python package MDAnalysis, which contains an implementation of the H-bond lifetime algorithm. MDAnalysis used the molecular trajectories from the production NVT simulations described above with an output frequency of 2 femtoseconds for a total of 20 picoseconds. The H-bond autocorrelation data were fit to an exponential function to determine the system's average time constant for H-bond lifetimes. The results are presented in Figure~\ref{fig:hbond_lifetime}. 

The macroscopic response to positive trends in temperature and pressure is the reduction and increase in liquid viscosity, respectively. As previously discussed, H-bond interactions are the main molecular scale contributor to macroscopic dynamic viscosity. Thus, longer H-bond lifetimes are associated with higher viscosity, while shorter H-bond lifetimes are associated with lower viscosity. Based on the results presented in Figure~\ref{fig:hbond_lifetime}, the EPM2 force field systems exhibited the molecular behavior expected from macroscopic trends in temperature and pressure in terms of H-bond lifetimes, while TraPPE and Zhang do not. This indicates the EPM2 force field to be a better model for the expected molecular behavior in the context of viscosity prediction of carbon dioxide hydrate systems at pre-nucleation conditions.

\subsubsection{Hydrogen Bond Density}
The hydrogen bond density of the simulated systems was calculated and is presented in Table~\ref{tbl:n_hbonds}. The H-bond density, \textit{n}, was calculated as the average number of hydrogen bonds over the simulated time normalized by the total number of molecules that may participate in hydrogen bonding in each simulated system. In the case of the simulations performed in this work, both water and carbon dioxide form hydrogen bonds, and thus the average number of hydrogen bonds over time was divided by the total number of molecules in the system. As previously discussed, the concentration of the molecular systems in this work is dictated by their thermodynamic state (temperature and pressure) and are described by Equation~\ref{eqn:KK}. As a result, simulations have a varying total number of molecules to achieve the required concentration, and thus the hydrogen bond density, \textit{n}, was normalized as described above to allow direct comparison between all simulated systems regardless of each system's total number of molecules. In these results, the hydrogen bond density of EPM2 systems decreased with increasing temperature. As previously discussed, hydrogen bonding and viscosity are directly related, additionally the expected macroscopic response to increased temperature is reduced viscosity. Thus, the decrease in hydrogen bond density with temperature is expected at the molecular level. This was only measured to be the case for the EPM2 systems (Table~\ref{tbl:n_hbonds}). Moreover, the macroscopic response to increased pressure is increased viscosity, and the EPM2 systems also demonstrated this behaviour as evidenced by the increasing trend in hydrogen bond density, \textit{n}, with pressure, while TraPPE and Zhang systems did not (Table~\ref{tbl:n_hbonds}). These observations offer further support to the H-bond lifetime analysis discussed above.

\begin{table}[H]
  \caption{The average number of hydrogen bonds per molecule over the simulated time for each temperature-pressure condition.}
  \label{tbl:n_hbonds}
  \newcolumntype{C}{>{\centering\arraybackslash}X}
\begin{tabularx}{\textwidth}{CCCCCC}
    \hline
\textbf{Temperature}	& \textbf{Pressure}	& \textbf{EPM2, n}	& \textbf{TraPPE, n}	& \textbf{Zhang, n}	& \textbf{Mean} \\
$^{\circ}$C & MPag & $\pm$0.0005 & $\pm$0.0005 & $\pm$0.0005 & $\pm$0.0005 \\
\hline
0	& 0	& 1.512	& 1.518	& 1.514	& 1.515	\\
4	& 0	& 1.264	& 1.505	& 1.520	& 1.430	\\ 
8	& 0	& 0.876	& 1.503	& 1.516	& 1.298	\\
\hline
0	& 2	& 1.760	& 1.488	& 1.491	& 1.579	\\
4	& 2	& 1.491	& 1.491	& 1.495	& 1.492	\\
8	& 2	& 1.025	& 1.492	& 1.484	& 1.334	\\
\hline
0	& 3	& 2.501	& 1.459	& 1.474	& 1.811	\\
4	& 3	& 2.111	& 1.468	& 1.478	& 1.685	\\
8	& 3	& 1.467	& 1.462	& 1.473	& 1.467	\\
\hline
\end{tabularx}
\noindent{\footnotesize{\textbf{Notes}: The average number of hydrogen bonds over the simulated time is calculated and divided by the total number of molecules that may participate in hydrogen bonding in each simulation: $\left ( n=\frac{N_{H-bonds}}{N_{molecules}} \right )$ for direct comparison across conditions and force fields.}}
\end{table}

\section{Conclusions and Future Work}
This work designed molecular simulations of carbon dioxide hydrate systems at pre-nucleation conditions through molecular dynamics. The TIP4P/Ice force field potential was used to model water molecules, while three commonly accepted force fields for carbon dioxide (EPM2, TraPPE, and Zhang) were examined for their performance in the context of transport property predictions of hydrate systems. Dynamic viscosity predicted by the molecular simulations was directly compared to previously collected experimental data to evaluate force field performance. Two formulations of viscosity were used - Green-Kubo and Stokes-Einstein. Generally, the Stokes-Einstein predictions suffered from high variation in predictions which likely stems from the unsuitability of the Stokes-Einstein assumptions for the systems simulated here. On average, the predicted viscosity for EPM2 systems were 61\% higher than experimental data, for TraPPE they were 65\% higher, and for Zhang they were 72\% higher. The EPM2 force field resulted in generally more accurate predictions of viscosity than other force fields. Diffusivity predictions were reported and their relationship viscosity was discussed. Experimental data for diffusivity is necessary for further validation of the simulation predictions.

This work conducted a hydrogen bond analysis as an attempt to examine possible molecular sources for the discrepancies between predicted dynamic viscosity and experimental data. The probability density functions of hydrogen bond length and angle were calculated to determine structural differences between the simulated systems. It was concluded that the molecular structure of hydrogen bonds did not appreciably change between simulations, indicating that hydrogen bond structure was not a likely source for the viscosity prediction discrepancies. The hydrogen bond lifetime analysis conducted in this work indicated that the EPM2 model exhibited trends in time constants with respect to temperature and pressure which were as expected by the relationship between hydrogen bond interaction and viscosity. Moreover, this result was supported by the normalized hydrogen bond density analysis, which indicated that the number of hydrogen bonds in EPM2 force field decreased with temperature while TraPPE and Zhang did not.

This work has attempted to produce molecular predictions of viscosity of carbon dioxide hydrate systems using currently available force fields and well-defined system design and equilibration procedure. We have reported the discrepancy between the predictions from the molecular models built and experimental data. In an attempt to begin searching for sources of error that may contribute to this discrepancy, this work has investigated the role of hydrogen bonding. Although the hydrogen bond analyses quantify some discrepancies between the carbon dioxide force fields investigated in the number of hydrogen and the extent of their interactions (lifetimes) further analyses should be performed in the future to investigate additional sources of error. Some potential sources of error to be investigated are the effects of water-carbon dioxide mixing on density, heat capacity, and freezing point. Additionally, the choice of mixing rules may be investigated. The result from such an analysis may present the opportunity for a re-parametrization of the carbon dioxide EPM2 force field for the context of transport property predictions of pre-nucleation gas hydrate systems to improve its prediction accuracy. A re-parametrized force field may prove useful to engineering applications in process design and control of carbon capture and sequestration technologies that make use of computational estimates of carbon dioxide hydrate slurry/suspension viscosity. This work can serve as a guide for future re-parametrization efforts, indicating the accuracy of current force field models, parameters to be adjusted and the baseline experimental data that should be used. Finally, the results presented in this work offer a quantitative characterization and comprehensive fundamental molecular scale analysis of pre-nucleation carbon dioxide hydrate systems and further our understanding of the material physics of this hydrate precursor material.

\bibliographystyle{abbrvnat}
\bibliography{main.bib}

\section*{Acknowledgments} 
The authors acknowledge the support from the Digital Research Alliance of Canada, Calcul Quebec, and WestGrid through computational resource grants, expertise, and technical support.

\section*{Funding}
Financial support for the work presented here was received from the Natural Sciences and Engineering Research Council of Canada (NSERC) through the Canada Graduate Scholarship Doctoral (CGS-D) award (A.G.), NSERC Discovery Grant number 206269 (P.S.), NSERC Discovery Grant number 206259 (M.M), NSERC Discovery Grant number 223086 (A.D.R.), Fonds de Recherche du Québec Nature et technologies (FRQNT) bourse de doctorat en recherche (S.M.), from the McGill Engineering Doctoral Award (MEDA) (A.G. and S.M.), and the James McGill Professorship (A.D.R.).

\clearpage
\beginsupplement

\section*{Supplementary Materials}
Here, we provide all supplementary materials used in our analysis. 

\end{document}